\documentclass[aps,preprint]{revtex4}

\begin{document}

\title{The role of symmetry in the short-time critical dymamics}

\author{T\^{a}nia Tom\'{e}}

\affiliation{Instituto de F\'{\i}sica\\
Universidade de S\~{a}o Paulo\\
Caixa Postal 66318\\
05315-970 S\~{a}o Paulo, S\~{a}o Paulo, Brazil}
\date{\today}

\begin{abstract}

We show that the short-time critical exponent $\theta$ related
to the critical initial slip in a stochastic model can be
determined by the time correlation of the order parameter.
In our procedure it suffices to start with 
an uncorrelated state with zero order parameter
instead of departing, as usually done, from an initial
state with a nonzero order parameter.
The proof uses the group of symmetry operations related to
the Markovian dynamics. Our scheme is extended to cover
models with absorbing states. 

PACS numbers: 05.70.Ln, 05.50.+q, 64.60.Ht

\end{abstract}

\maketitle

\section{Introduction}

The universal behavior occurring
in the first steps of a Monte Carlo simulation,
the short-time dynamics, has
been amply investigated \cite{jan,huse,li0,li,zheng,zheng1,jaster} 
in the last years.
According to renormalization group arguments \cite{jan} the early time
behavior of the order parameter (the magnetization, for example, in the case
of the Ising model) follows a power law with a critical universal
exponent $\theta $. 
The numerical calculation of the exponent $\theta $ is performed
by placing the system at the critical point and departing from a
configuration where the order parameter $m_0$ is very small.

The purpose of this article is to show that it is possible to determine the
exponent $\theta $ by starting with a configuration in which the order
parameter is identically zero. 
In our approach we do not measure the order parameter itself,
which is zero, but its time coorelation.
If $M(t)$ denotes de instantaneous order
parameter we show that the quantity 
\begin{equation}
Q(t)=\langle M(t)M(0)\rangle
\label{1}
\end{equation}
follows a power law 
\begin{equation}
Q(t)\sim t^\theta
\end{equation}
where the initial configuration is uncorrelated and such that 
\begin{equation}
\langle M(0)\rangle =0
\end{equation}
This result is general and can be applied to any lattice system described by
a Markovian process and such that the transition probability is
invariant under a given group of
symmetry operations. In this procedure, which was previously deduced for
models with up-down symmetry \cite{mario}.
Here we generalize the scheme to inclued models with
other types of symmetries. In particular we apply the method 
to systems with antiferromagnetic ordering and to
the Baxter-Wu model in a triangular lattice \cite{baxter}. 
Employing this procedure it was
recently calculated by numerical simulations \cite{everaldo}, with an
excellent precision, the exponent $\theta $ associated to the Baxter-Wu model
with triplet interactions in a triangular lattice.

We consider also the short-time behavior of intrinsic irreversible
models with absorbing states,
such as the contact process. In this
case, we demonstrate here that it suffices 
to depart from a configuration with
just one occupied site instead of starting from a configuration where the
density of particles is finite and small. Therefore, the short-time behavior
of the order parameter is identical to the behavior found by time-dependent
simulations \cite{iwan,marro} departing from a unique initial seed.
In both methods it is necessary to place
the system in its critical point. 

\section{Transition probability and symmetry}

In this section and in the following we develop the formalism leading
to the expression (\ref{1}). For this purpose we
consider the class of Markovian processes defined on a lattice whose
probability distribution $P(\sigma,t)$ satisfies the equation 
\begin{equation}
P(\sigma,t)=\sum_{\sigma^{\prime }}T(\sigma,\sigma^{\prime },t)
P_0(\sigma^{\prime })
\end{equation}
where $T(\sigma,\sigma^{\prime },t)$ is the probability of reaching the
configuration $\sigma$ from configuration $\sigma^{\prime }$ in an interval of 
time $t$ and $P_0(\sigma^{\prime })$ is the initial probability 
distribution. We use the
notation $\sigma=(\sigma_1,\sigma_2,...,\sigma_N)$ where $N$ is the 
number of sites in the
lattice and $\sigma_i$ is the random variable attached to the $i$-th
site and that takes two values.

If the system evolves in time according to a
master equation (continuous time Markovian process) 
\begin{equation}
\frac d{dt}P(\sigma,t)=\sum_{\sigma^{\prime }}\{W(\sigma,\sigma^{\prime })
P(\sigma^{\prime},t)-W(\sigma^{\prime },\sigma)P(\sigma,t)\}
\end{equation}
then
the transition probability $T(\sigma,\sigma^{\prime },t)$ are 
the elements of the
matrix $T$ given by 
\begin{equation}
\widehat{T}=\exp \{t\widehat{W}\}
\end{equation}
where $\widehat{W}$ is the matrix whose elements are 
\begin{equation}
\widehat{W}(\sigma^{\prime },\sigma)=W(\sigma^{\prime },\sigma)
\qquad \sigma^{\prime }\neq \sigma
\end{equation}
and 
\begin{equation}
\widehat{W}(\sigma,\sigma)=-\sum_{\sigma^{\prime }(\neq \sigma)}
W(\sigma^{\prime },\sigma)
\end{equation}

Let $R$ be a symmetry operation that leaves the transition probability
invariant, or, equivalently, the matrix $W$ invariant, that is 
\begin{equation}
W(R\sigma,R\sigma^{\prime })=W(\sigma,\sigma^{\prime })
\end{equation}
and by consequence
\begin{equation}
T(R\sigma,R\sigma^{\prime },t)=T(\sigma,\sigma^{\prime },t)
\label{7}
\end{equation}
For simplicity, we will consider here only models in which the symmetry
operation $R$ changes the sign of the order parameter, that is, 
\begin{equation}
M(R\sigma)=-M(\sigma)
\label{8}
\end{equation}
and that $M(\sigma)$ is linear in $\sigma$, that is,
\begin{equation}
M(\sigma)=\sum_i \mu_i \sigma_i
\label{11}
\end{equation}

\section{Time-dependent behavior}

We will focus on the time-dependent behavior of the average 
\begin{equation}
\langle M(\sigma)\rangle _t=\sum_\sigma M(\sigma)P(\sigma,t)
\end{equation}
of the order parameter $M(\sigma)$. Its time evolution is given by 
\begin{equation}
\langle M(\sigma)\rangle _t=\sum_\sigma\sum_{\sigma^{\prime }}M(\sigma)
T(\sigma,\sigma^{\prime
},t)P_0(\sigma^{\prime })
\label{12}
\end{equation}
where the initial state $P_0(\sigma)$ is uncorrelated with a nonzero
magnetization. That is, the initial magnetization 
\begin{equation}
\langle M(\sigma)\rangle _0=\sum_\sigma M(\sigma)P_0(\sigma)=Nm_0
\end{equation}
is nonzero, where $N$ is the number of sites of the lattice and $m_0$ is a
small quantity. As stated by the short-time scaling theory, 
the order parameter
folows, at the critical point, a power law behavior 
\begin{equation}
\langle M(\sigma)\rangle _t\sim m_0 t^\theta
\label{16} 
\end{equation}
for small values of $m_0$. 
According to this theory yet the
initial state is prepared in such a way that all sites are
uncorrelated with a nonzero (and small) initial order parameter $m_0$.
In order to set up such an initial state, one attributes to
each site a magnetization $m_i=m_0 \mu_i$. Or equivalentely,
the spin of the $i$-th site will be $\sigma_i=\mu_i$ with probability
$(1+m_0)/2$ and will be $\sigma_i=-\mu$ with probability
$(1-m_0)/2$. The initial probability $P_0(\sigma)$ 
can then be written as
\begin{equation}
P_0(\sigma)= \Phi_0 \prod_i \{ 1+m_0 \mu_i \sigma_i \}
\label{17}
\end{equation}
where
\begin{equation}
\Phi _0=\frac 1{2^N}
\end{equation}
Notice that 
using equation (\ref{11}) and (\ref{17}) we can trivially find that
\begin{equation}
\langle M(\sigma)\rangle _0=
\sum_\sigma M(\sigma)P_0(\sigma)= N a m_0
\end{equation}
where $a$ is the constant
\begin{equation}
a=\frac1N \sum_j [\mu_j]^2
\end{equation}

For small values of $m_0$, the expansion of the initial probability
$P_0(\sigma)$ in powers of $m_0$ gives, up to
linear terms in $m_0$ the following expression
\begin{equation}
P_0(\sigma)=\Phi _0\{1+m_0 M(\sigma) \}
\end{equation}
Substituting this expression in equation (\ref{12}) we get 
\[
\langle M(\sigma)\rangle _t=\sum_\sigma\sum_{\sigma^{\prime }}M(\sigma)
T(\sigma,\sigma^{\prime },t)\Phi _0+
\]
\begin{equation}
+\sum_\sigma\sum_{\sigma^{\prime }}M(\sigma)T(\sigma,\sigma^{\prime},t)
\Phi _0 m_0 M(\sigma^{\prime })
\label{24}
\end{equation}

Now, the first term on the right hand side vanishes 
identically due to the following reasoning. Since the symmetry
operation $R$ leaves the transition probability invariant 
but changes the sign of the order parameter, we have
\begin{equation}
\sum_\sigma\sum_{\sigma^{\prime }}M(R\sigma)T(R\sigma,R\sigma^{\prime},t)
\Phi _0 =
- \sum_\sigma\sum_{\sigma^{\prime }}M(\sigma)T(\sigma,\sigma^{\prime},t)
\Phi _0
\end{equation}
By a change of variable, $R\sigma\to \sigma$, the left-hand side of this
equation equals the first term of the right-hand side
of equation (\ref{24}) so that it vanishes.
Therefore 
\begin{equation}
Q(t)=\lim_{m_0\rightarrow 0}\frac{\langle M(\sigma)\rangle _t}{m_0}=
\sum_\sigma\sum_{\sigma^{\prime }}M(\sigma)T(\sigma,\sigma^{\prime },t)
M(\sigma^{\prime })\Phi _0
\end{equation}
and, from equation (\ref{16}) it follows that
\begin{equation}
Q(t) \sim t^\theta
\end{equation}

\section{Applications}

\subsection{Models with up-down symmetry}

We begin with a simple example, namely the 
ferromagnetic Ising model coupled to a
stochastic dynamics such as the Metropolis
algorithm. The order parameter is defined by 
\begin{equation}
M(\sigma )=\sum_i\sigma _i
\end{equation}
where the summation is over all sites of the lattice. 
For the present case $\mu_i=+1$ for all sites 
of the lattice. Here the symmetry operation $R$,
with the properties given by equations (\ref{7})
and (\ref{8}), is the one in which the
up-down symmetry is observed, that is,
the operation that changes the sign of each spin
variable $\sigma_i \to -\sigma_i$.

The short-time behavior for the Ising model
has been already studied through the present formalism 
\cite{mario}. Besides, using this formalism, it has
been possible to determine the short-time behavior for
several irreversible models (i. e., lacking detailed balance) 
\cite{mario,td,tania} with up-down symmetry dynamics. These include
for instance the majority vote model and similar nonequilibrium
models \cite{mario,mendes}.

\subsection{Models with antiferromagnet ordering}

In this case the system is divided into two sublattices 
$A$ and $B$. The order parameter is defined by 
\begin{equation}
M(\sigma)= \sum_{i \in A} \sigma_i - \sum_{i \in B} \sigma_i
\end{equation}
Therefore, for this case one has $\mu_i=+1$ if $i \in A$
and $\mu_i=-1$ if $i \in B$.
The symmetry operation $R$ is a translation
such that a given site of one sublattice
goes into a site of the other sublattice.

\subsection{Baxter-Wu model}

We consider in this subsection the Baxter-Wu 
model with triplet interactions defined on a
triangular lattice \cite{baxter,everaldo}. The lattice is composed
of three sublattices which we denote by $A$, $B$, and $C$. 
The Baxter-Wu model does not have a global symmetry
but semi-global symmetries. The Hamiltonian of the model 
and a fortiori the transition probability
is invariant if we change the signs 
of two sublattices leaving the third invariant.
It is convenient therefore to define three symmetry 
operations, denoted by $R_A$, $R_B$ and $R_C$.
The symmetry operation $R_A$
changes the signs of the spins belonging
to the sublattices $B$ and $C$ and leaves 
the signs of the spins of sublattice $A$
invariant. Similar definitions can be stated
for $R_B$ and $R_C$. Each of these symmetry operations
leaves the Baxter-Wu Hamiltonian invariant and 
a fortiori the transition probability invariant.

We take as the order parameter the magnetization 
of one of the sublattices, say, sublattice $A$,
given by 
\begin{equation}
M_A(\sigma )=\sum_{i\in A}\sigma _i
\end{equation}
Comparing it with equation (\ref{11}) we have that
$\mu_i=1$ if $i\in A$ and $\mu_i=0$ if $i\in B$ or
$i\in C$. The symmetry operation $R_B$
(or $R_C$) changes the sign of $M_A(\sigma)$ and
leaves the transition probability invariant. 
According to the formalism developed in 
the previous section we conclude that
\begin{equation}
Q_A(t)=\lim_{m_0\rightarrow 0}\frac{\langle M_A(\sigma)\rangle _t}{m_0}=
\sum_\sigma\sum_{\sigma^{\prime }}M_A(\sigma)T(\sigma,\sigma^{\prime },t)
M_A(\sigma^{\prime })\Phi _0
\end{equation}
will behave as 
\begin{equation}
Q_A(t) \sim t^\theta
\end{equation}
Equivalently, we may demonstrate that
the analogous quantities $Q_B(t)$ and $Q_C(t)$
related to the magnetizations $M_B(\sigma)$ and $M_C(\sigma)$ 
of sublattices $B$ and $R_C$,
respectively, will behave as $t^\theta$.

We may also use as the order parameter the total magnetization
\begin{equation}
M(\sigma)= \sum_i \sigma_i
\end{equation}
which we write as the sum of the magnetizations
of the three sublattices
\begin{equation}
M(\sigma) = M_A(\sigma)+M_B(\sigma)+M_C(\sigma)
\label{35}
\end{equation}
which leads to 
\begin{equation}
Q(t)=
\sum_\sigma\sum_{\sigma^{\prime }}M(\sigma)T(\sigma,\sigma^{\prime},t)
M(\sigma^{\prime })\Phi _0
\label{36}
\end{equation}
Substituting (\ref{35}) into (\ref{36}) we see that $Q(t)$ is
a sum of nine terms. The terms that involve magnetizations
of distinct sublattices will vanish. For instance, the term
that involve $M_A$ and $M_B$ will change sign by the use
o the symetry operation $R_A$. The nonvanishing terms are the
three terms that involve the same magnetization. One
concludes that the quantity 
\begin{equation}
Q(t)=Q_A(t)+Q_B(t)+Q_C(t)
\end{equation}
and therefore will behave as $t^\theta$.
This procedure was used \cite{everaldo} to determine the 
exponent $\theta $. The numerical results give very precise values
for the exponent when compared with the results coming 
from simulations with nonzero initial magnetization.

\section{Contact model}

Now we discuss the short-time behavior fo models with an 
absorbing state.
These models do not possess symmetry operations 
like the ones defined in the preceeding sections.
Due to this fundamental difference we need to proceed by 
introducing another approach. 
The simplest example of this type of
model is the contact process \cite{marro}.
Such model is defined in a lattice
and each microscopic state is identified with 
$\sigma =(\sigma _1,\sigma_2,...,\sigma _N)$ 
where $\sigma _i=0$ or $1$ according wether the site $i$ is
empty or occupied by a particle. It evolves in time according to local
Markovian rules where particles are catalitically created
and spontaneously anihilated.

The initial probability is such that all
sites are uncorrelated and such that the average $\langle \sigma _i\rangle
=\rho _0$, that is, 
\begin{equation}
P_0(\sigma )=\prod_i\{a(1-\sigma _i)+b\sigma _i\}
\end{equation}
where 
\begin{equation}
a=1-\rho _0\qquad \qquad b=\rho _0
\end{equation}
is the total number of sites in the lattice. 

Following the short-time scaling theory
the order parameter $\langle n(\sigma) \rangle$
given by
\begin{equation}
\langle n(\sigma )\rangle _t=\sum_{\sigma ^{\prime }}\sum_\sigma 
n(\sigma )T(\sigma,\sigma ^{\prime },t)P_0(\sigma ^{\prime })
\end{equation}
where 
\begin{equation}
n(\sigma )=\sum_i\sigma _i
\end{equation}
is the number of particle,
behaves, at the critical point, as
\begin{equation}
\langle n(\sigma) \rangle \sim \rho_0 t^\theta 
\end{equation}
Consequently the quantity 
\begin{equation}
Q(t)=\frac 1N\lim_{\rho _0\rightarrow 0}
\frac{\langle n(\sigma )\rangle _t}{\rho _0}
\end{equation}
has a similar behavior in the early time regime 
\begin{equation}
Q(t) \sim t^\theta
\end{equation}

For small values $\rho _0$ of we have 
\begin{equation}
P_0(\sigma )=a^N\Phi _0(\sigma )+a^{N-1}b\sum_j\Phi _j(\sigma )
\end{equation}
where 
\begin{equation}
\Phi _0(\sigma )=\prod_i(1-\sigma _i)
\end{equation}
is the probability distribution such that the configuration $(0,0,0,...,0)$
(all sites empty) has probability $1$ and the other configurations have zero
probability, 
\begin{equation}
\Phi _j(\sigma )=\sigma _j\prod_{i(\neq j)}(1-\sigma _i)
\end{equation}
is the probability distribution such that the configuration 
$(0,0,...,1,...0) $ (a particle placed at the $j$-th site 
and all other sites empty) has probability $1$ and all 
other configurations have zero probability.

The average $\langle n(\sigma )\rangle _t$ can then be 
written as a sum of two parts 
\[
\langle n(\sigma )\rangle _t=
a^N\sum_{\sigma ^{\prime }}\sum_\sigma n(\sigma )T(\sigma,\sigma ^{\prime },t)
\Phi _0(\sigma ^{\prime })+ 
\]
\begin{equation}
+a^{N-1}b\sum_j\sum_{\sigma ^{\prime }}\sum_\sigma n(\sigma )
T(\sigma ,\sigma ^{\prime},t)\Phi _j(\sigma ^{\prime })
\end{equation}
Since the contact process has an abosrbing state devoided of particles,
the first term vanishes identically because $\Phi _0(\sigma )$ is the
absorbing state. Therefore, using the translational invariance we obtain 
\begin{equation}
Q(t)=\sum_{\sigma ^{\prime }}\sum_\sigma n(\sigma )
T(\sigma ,\sigma ^{\prime },t)\Phi_j(\sigma ^{\prime })
\end{equation}
To calculate numerically $Q(t)$, we start from a configuration with just one
occupied site and determine the number of occupied sites at time $t$.

According to the scaling relations established for the time-dependent
behavior of the contact model in which the simulation
is started with just one occupied site, the average number
of particles $n_p(t)$ behaves as \cite{iwan,marro}
\begin{equation}
n_p(t) \sim t^\eta
\end{equation}
As $Q(t)$ is identified with $n_p(t)$ 
so the exponent $\theta $ is identified with the exponent $\eta $
\cite{hinri}.

Let us now calculate the time correlation of a given site, say the site $j$.
It is given by 
\begin{equation}
A(t)=\sum_{\sigma ^{\prime }}\sum_\sigma \sigma _j
T(\sigma ,\sigma ^{\prime },t)\sigma_j^{\prime }P_0(\sigma ^{\prime })
\end{equation}
and behaves, according to the short-time scaling theory, as
\begin{equation}
A(t) \sim t^\lambda
\end{equation}

Now
\begin{equation}
\sigma _jP_0(\sigma )=b\sigma _j\prod_{i(\neq j)}\{a(1-\sigma _i)+b\sigma _i\}
\end{equation}
where we have used the obvious relations 
$\sigma _j(1-\sigma _j)=0$ and $\sigma_j\sigma _j=\sigma _j$. 
Therefore, in the limit $\rho _0\rightarrow 0$, we get 
\begin{equation}
\lim_{\rho _0\rightarrow 0}\frac 1{\rho _0}\sigma _jP_0(\sigma )=
\Phi _j(\sigma )
\end{equation}
Consequently, 
\begin{equation}
B(t)=\lim_{\rho _0\rightarrow 0}\frac{A(t)}{\rho _0}=
\sum_{\sigma ^{\prime}}\sum_\sigma \sigma _jT(\sigma ,\sigma ^{\prime},t)
\Phi _j(\sigma ^{\prime })
\end{equation}
so that $B(t)$ behaves as
\begin{equation}
B(t) \sim t^\lambda
\end{equation}
Given that the initial particle seed is placed at
a given site, the quantity $B(t)$ is the probability 
that this site be occupied at time $t$.

The exponent $\lambda$ is related to dynamic exponent $z$
by $\lambda=d/z-\theta$ \cite{mario}.
Since the exponent $\theta$ was identified as the exponent $\eta$, it 
follows that $\lambda=d/z-\eta$. Now, from the hyperscaling relation
for the contact process we have $d/z-\eta=2\delta$ \cite{marro}
where $\delta$ is the exponent associated to the survival 
probability. Therefore it follows that $\lambda=2\delta$.

>From the formalism developed here we conclude that
the study of the short-time behavior of the contact process, 
discussed in reference \cite{hinri}, is as a matter
of fact equivalent to the study of this model by means of
the time-dependent technique. 
Moreover, the critical exponents associated to the short
time dynamics for the contact model, as well as the
relation among them, are equivalent to those found
for the time-dependent simulations.

\section{Conclusion}

We have shown that the short-time critical exponent $\theta$
of several models invariant under a given 
group of symmetry
can be calculated from the time correlation
of the order parameter
\begin{equation}
Q(t)=\langle M(t)M(0)\rangle
\end{equation}
where $\langle f(t)g(0)\rangle $ is a notation defined by 
\begin{equation}
\langle f(t)g(0)\rangle =\sum_\sigma \sum_{\sigma ^{\prime }}f(\sigma
)T(\sigma ,\sigma ^{\prime },t)g(\sigma ^{\prime })\Phi _0
\end{equation}
We  have also obtained similar formula for the contact process
and shown that the short-time critical exponent $\theta$
is equal to time-dependent critical exponent $\eta$.
Finally, the results obtained here for the continuous time Markovian
processes can be straitforwardly extended to the probabilistic 
cellular automata (discrete time Markovian process). 

\begin{center}
{\bf Acknowledgements}
\end{center}
I am very grateful to J. R. Drugowich de Fel\'{\i}cio
for useful and instigating discussions about this
issue. This work was suported in part by the Brazilian
agency CNPq.

\end{document}